\documentclass[fleqn,usenatbib]{mnras}

\usepackage[T1]{fontenc}

\DeclareRobustCommand{\VAN}[3]{#2}
\let\VANthebibliography\thebibliography
\def\thebibliography{\DeclareRobustCommand{\VAN}[3]{##3}\VANthebibliography}

\usepackage{graphicx}	
\usepackage{amsmath}	
\usepackage{amssymb}	

\usepackage{newtxtext,newtxmath}

\title[Linearity]{Linearity: galaxy formation encounters an unanticipated empirical relation}

\author[Stephen Lovas]{Stephen Lovas$^{1}$%
\thanks{Contact e-mail: \href{mailto:stephen.lovas@louisville.edu}{stephen.lovas@louisville.edu}}%
\\
$^{1}$University of Louisville, 
2301 South Third Street, 
Louisville, KY 40292, USA}

\date{Accepted XXX. Received YYY; in original form ZZZ}

\pubyear{2022}

\begin{document}
\label{firstpage}
\pagerange{\pageref{firstpage}--\pageref{lastpage}}
\maketitle

\begin{abstract}
Measurements from galaxies spanning a broad range of morphology reveal a linear scaling of enclosed dark to luminous mass that is not anticipated by standard galaxy formation cosmology. The linear scaling is found to extend from the inner galactic region to the outermost data point. Uncertainties in the linear relation are narrow, with rms = 0.31 and $\sigma$ = 0.31. It is unclear what would produce this linearity of enclosed dark to luminous mass. Baryonic processes are challenged to account for the linear scaling, and no dark matter candidate possesses a property that would result in a linear relation. The linear scaling may indicate new dark matter candidates, or an astrophysical process beyond standard galaxy formation theory.
\end{abstract}




\section{Introduction}

According to standard cosmological theory, galaxies formed in the early universe within the potential wells of dark matter haloes \citep{White1978, Blumenthal1984, Freeman2003, Bromm2009}. In a process of hierarchical galaxy formation, small haloes merged with larger haloes to form the galaxies, groups, and clusters observed in the present universe.

In standard cosmology, dark matter haloes are comprised of theoretical particles that neither emit, nor absorb, nor reflect radiation at any wavelength. Well motivated dark matter candidates include Weekly Interacting Massive Particles \citep[WIMPs;][]{Iwanus2019,Cook2020,Basu2021,Thorpe-Morgan2021}, warm dark matter \citep{Chatterjee2019,Gilman2020,Garzilli2021,Hermans2021}, self-interacting dark matter \citep{Despali2019,Robles2019,Robertson2021,Correa2021}, axions \citep{Leong2019,Galanti2020,Bauer2021,Giare2021}, and primordial black holes \citep{Carr2018,Stegmann2020,Kim2021,Sureda2021}.

Discrepancies, however, have been noted between dark matter simulations and observations \citep[see][for a review]{bullock2017}. In contrast to cuspy cores predicted by dark matter models \citep{navarro1996,moore1998}, nearly constant density cores are found in a wide range of galaxies \citep[e.g.][]{binney2001,keeton2001,deblok2001,gentile2007,spano2008,bekki2009,amorisco2013,rodrigues2017}.  

Dark matter simulations also predict the majority of the most massive subhaloes of the Milky Way are too dense to host any of its bright satellites \citep{Boylan-Kolchin2011,dicintio2011,garrison-kimmel2014}. 

\begin{figure}
\begin{center}
\includegraphics[width=8cm]{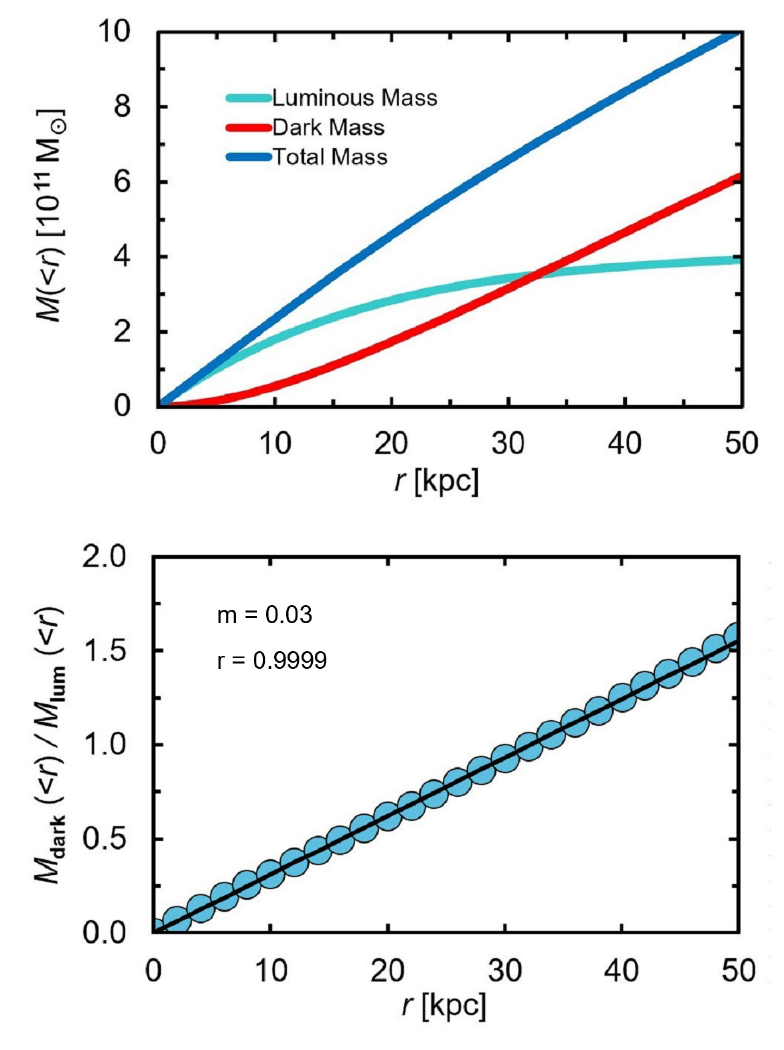}
\end{center}
\textbf{\caption{Enclosed mass of pressure supported E0 galaxy NGC4636. Upper panel plots the profiles of luminous, dark, and total mass. Velocities of 174 globular clusters are used as dynamical tracers. Lower panel shows the ratio of NGC4636 enclosed dark mass to luminous mass as a function of radius. The ratio of enclosed dark to luminous mass scales linearly with radius. Solid black line delineates the slope of the linear relation. The slope value \textit{m} is noted in the upper left corner of the panel together with the correlation coefficient r. Slope is derived using the least squares method with intercept taken to be zero.}} \label{fig:dmfig1}
\end{figure}

In addition to tensions between dark matter simulations and observations, an empirical relation between dark and luminous mass has been suggested through analysis of a small number of galaxies \citep{Lovas2014} including elliptical galaxy NGC4636. Enclosed total mass of NGC4636 increases continuously to the outermost data point even as its luminous mass converges toward a limiting mass \citep[Fig.~\ref{fig:dmfig1} upper panel;][]{Schuberth2006}. The authors do not quantify the uncertainties since they are mostly of a systematic nature. As total enclosed mass of NGC4636 increases continuously, the ratio of enclosed dark to luminous mass scales linearly with radius. The linear scaling reaches from the inner galactic region to the outermost data point (Fig.~\ref{fig:dmfig1} lower panel). It is unclear what would produce this close linear relation. No dark matter candidate possesses a theoretical property that would lead to a linear scaling. 

To ameliorate dark matter tensions, modifications have been proposed to the laws of gravity, notably MOND \citep{Milgrom1983}. MOND, however, is in strong disagreement with gravitational lensing measurements \citep{Clowe2006} and conflicts with observations on scales ranging from dwarf galaxies \citep{Sanchez-Salcedo2013} to galaxy clusters \citep{Silk2005}.

The core-cusp and too-dense subhaloe discrepancies could be mitigated through baryonic processes such as star formation and supernova feedback \citep{governato2012,zolotov2012,madau2014,chan2015,wetzel2016,dashyan2018}. Baryonic processes, however, are challenged to account for the linear scaling of enclosed dark to luminous mass. The profile of dark and luminous mass is unique to each galaxy due to the galaxy's  individual size, morphology, and quantity of dark and luminous mass. To produce a linear scaling of enclosed dark to luminous mass, star formation and supernova feedback processes would require individual tuning to each galaxy.

It would be of interest to determine if the linear scaling of enclosed dark to luminous mass exhibited by pressure supported galaxy NGC4636 would be found also in a sample of rotationally supported galaxies. 

\section{Linearity of rotationally supported galaxies} \label{rotationally}

To examine whether a linear scaling of enclosed dark to luminous mass might also be found in rotationally supported galaxies, we sample four galaxies from the Spitzer Photometry \& Accurate Rotation Curves (SPARC) catalogue of 175 galaxies \citep{Lelli2016}. 

\begin{figure*}
\begin{center}
\includegraphics[width=18cm]{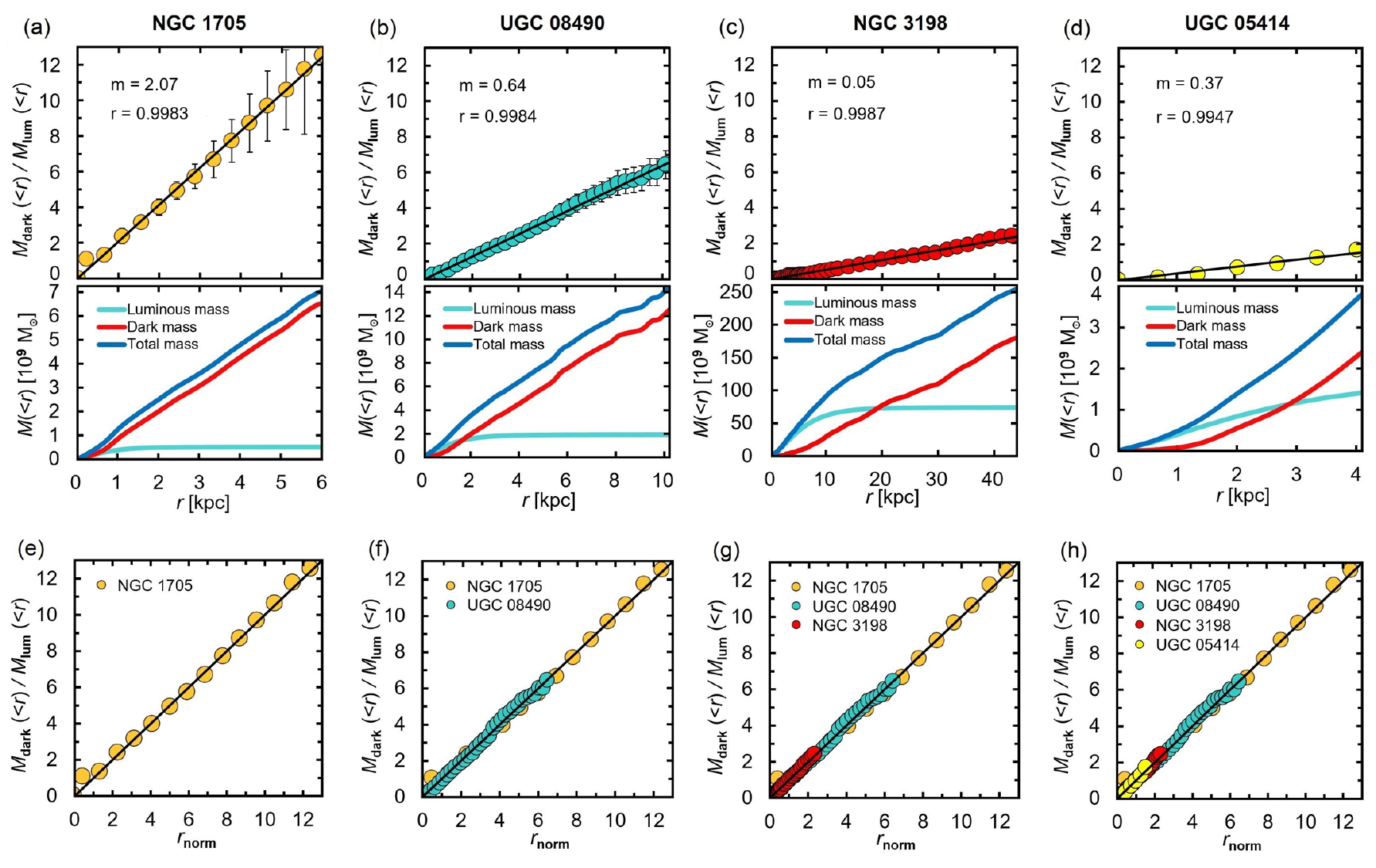}
\end{center}
\caption{Enclosed mass of four galaxies from the SPARC catalogue. Upper section of panels (a)-(d) displays the ratio of enclosed dark to luminous mass as a function of radius. With each galaxy the ratio follows a linear relation. Solid lines delineate the slope of the relation. The slope value is noted in the upper left corner together with the correlation coefficient. Error bars show the uncertainty in the ratio of enclosed dark to luminous mass. Error bars smaller than the size of the points are not visible. Lower section of panels (a)-(d) plot the profiles of luminous, dark, and total mass for the four galaxies. The linear scaling of enclosed dark to luminous mass as a function of radius is more easily observed. In panels (e)-(h), the ratio of enclosed dark to luminous mass is re-plotted with radii normalised. Radii are normalised by applying slope \textit{m} to \textit{r}, such that  \textit{m}\textit{r} = $\textit{r}_{norm}$. From left to right, data points are over-plotted upon the previous panel. The solid black line identifies a slope of unity.} 
\label{fig:dmfig2}
\end{figure*}

 Figure~\ref{fig:dmfig2} plots enclosed mass of the four galaxies. The upper section of panels (a)-(d) displays the ratio of enclosed dark to luminous mass as a function of radius. With all four galaxies, the ratio of enclosed dark to luminous mass follows a linear relation. Solid black lines delineate the slope of that relation. The slope value \textit{m} is noted in the upper left corner together with the correlation coefficient r.

The lower section of panels (a)-(d) plots the profiles of luminous, dark, and total mass for the four galaxies. The linear scaling of enclosed dark to luminous mass as a function of radius is more readily observed. For NGC1705 and UGC05414, panels (a) and (d) respectively, the mass profiles are dissimilar. Never the less, the ratio of enclosed dark to luminous mass scales as \textit{r} in both galaxies. 

By contrast, the mass profiles are similar for UGC05414, Fig.~\ref{fig:dmfig2}(d), and for NGC4636, Fig.~\ref{fig:dmfig1}. NGC4636, however, is a pressure supported elliptical of E0 morphology, whereas UGC05414 is a rotationally supported irregular of IABm morphology. The radius of NGC4636 exceeds that of UGC05414 by a factor of 12, and its mass exceeds that of UGC05414 by a factor ${>}$ 2 dex. None the less, for both galaxies, the radial scaling of enclosed dark to luminous mass follows a linear relation.

Consistent with NGC4636 (Fig.~\ref{fig:dmfig1}), in each galaxy of Fig.~\ref{fig:dmfig2}, the radial mass profile of dark matter scales with luminous mass to produce a linear scaling of enclosed dark to luminous mass. With each galaxy the linear scaling extends from the inner region to the last kinematic data point.

To place the galaxies of Fig.~\ref{fig:dmfig2} on a uniform scale, their radii are normalised by applying slope \textit{m} to \textit{r}, such that  \textit{m}\textit{r} = \textit{r}$_{norm}$.

With radii normalised, the ratio of enclosed dark to luminous mass is re-plotted in panels (e)-(h). From left to right the data points are over-plotted upon the previous panel. The ratio of enclosed dark to luminous mass for the four galaxies scales proportionally with radius. With radius normalised, the slope of proportionality is unity, solid black line, panels (e)-(h). 

Four galaxies from the SPARC catalogue exhibit in Fig.~\ref{fig:dmfig2} a linear scaling of enclosed dark to luminous mass. To identify if such linearity would extend to the broad database, we evaluate a large sample of galaxies from the catalogue.

\subsection{Analysis of the SPARC data} \label{sec:analysis}

The SPARC catalogue of 175 galaxies uses near-infrared (3.6 \micron) surface photometry from NASA's Spitzer space telescope to trace the distribution of stellar mass. Galaxies in the catalogue range in morphology from S0 to Sd to Irr. H$\alpha$ and H {\footnotesize I} measurements from 54 published sources are incorporated into the database. Numerous researchers have drawn upon the catalogue for their work. \citep[e.g.][]{Bernal2018,Santos2018,Naik2019,Tortora2019,Ponomareva2021}. The complete database is located at \url{http://astroweb.cwru.edu/SPARC}. 

 The compilers of the database describe in detail the methods they employed to determine stellar, gas, and observed rotation velocities through the application of 3.6 $\micron$ surface brightness profiles, H {\footnotesize I} surface density profiles, and H {\footnotesize I} / H$\alpha$ rotation curves \citep{Lelli2016}. SPARC data on the stellar components are for stellar mass-to-light ratio $\Upsilon_\ast$ = 1 to facilitate conversion to other mass-to-light ratios. When a bulge is present, the stellar rotation velocity is further decomposed into bulge and disk. A bulge is present in 32 of the SPARC galaxies. For the gas contribution, a thin disk is assumed. SPARC data adjust H {\footnotesize I} measurements by a factor 1.33 to account for helium. Increasing the H {\footnotesize I} factor to 1.4 to account for both helium and metals produces no impact on the results presented here.

\cite{Gentile2004} determine that luminous matter of several galaxies now in the SPARC catalogue is distributed as in a thin disk. \cite{Hallenbeck2014} find that for other galaxies now in the SPARC database, varying the thin stellar disk to allow for finite thickness has little or no effect. For consistency, the present paper adopts the thin disk geometry. Beyond $\textit{r}_{norm}$ = 1 spherical dark mass is the dominant mass component. Dark mass is ascribed to the amplitude of rotation velocity unaccounted for by the luminous components, where 
\begin{equation}\label{eq_1}
    V^2_{obs}(\textit{r}) = V^2_{lum}(\textit{r}) + V^2_{dark}(\textit{r}).
\end{equation}

In Equation~(\ref{eq_1}) velocity of the luminous contribution encompasses three components: the stellar disk $V_{disk}$, stellar bulge $V_{bul}$, and gas disk $V_{gas}$, such that 
\begin{equation}\label{eq_2}
    V^2_{lum}(\textit{r}) = \Upsilon_\ast V^2_{disk}(\textit{r}) + \Upsilon_\ast V^2_{bul}(\textit{r}) + V^2_{gas}(\textit{r}),  
\end{equation}
where $\Upsilon_\ast$ is the stellar mass-to-light ratio.

With 48 of the 175 SPARC galaxies, $V_{gas}$ is negative in the inner regions. This occurs when the gas distribution has a central depression and material in the outer areas exerts a stronger gravitational force than in the inner regions \citep{Lelli2016}. For those data points with negative $V_{gas}$, velocity of the luminous contribution takes the form 
\begin{equation}\label{eq_3}
    V^2_{lum}(\textit{r}) = \Upsilon_\ast V^2_{disk}(\textit{r}) + \Upsilon_\ast V^2_{bul}(\textit{r}) - V^2_{gas}(\textit{r}).  
\end{equation}
$V_{gas}$ is negative for 361 of the 3391 data points in the SPARC database. 

\cite{Meidt2014} report that at 3.6 $\micron$ a single $\Upsilon_\ast$ = 0.6 $\textit{M}_\odot$/$\textit{L}_\odot$ is feasible for a Chabrier initial mass function. The present study adopts a single $\Upsilon_\ast$ = 0.6 $\textit{M}_\odot$/$\textit{L}_\odot$ at 3.6 $\micron$ for all galaxies in the SPARC catalogue. The value $\Upsilon_\ast$ = 0.6 $\textit{M}_\odot$/$\textit{L}_\odot$ is applied uniformly to both the stellar disk and stellar bulge. A single $\Upsilon_\ast$ = 0.6 $\textit{M}_\odot$/$\textit{L}_\odot$ results in negative dark mass at 244 of the 3391 data points. To preclude a negative ratio of dark to luminous mass, negative dark mass is assigned a value of zero in the analysis. If negative dark mass were permitted, $\sigma$ would increase 0.01 and rms would increase 0.01.

\subsection{Selection of data from the SPARC catalogue} \label{sec:data}

Typical of an extensive database, measurements within the SPARC catalogue vary in certainty. To varying degrees, for example, the galaxies in the catalogue display asymmetries, non-circular motions, and off-sets between H {\footnotesize I} and stellar distributions \citep{Lelli2016}. Twelve such galaxies are identified by \cite{Lelli2016} to be of low quality. If these 12 galaxies are removed from the database, $\sigma$ increases 0.004 and rms increases 0.004.

   Accuracy of H {\footnotesize I} rotation curve measurements diminishes in nearly face-on galaxies, where inclination \textit{i} $< 40\degr$ \citep{Kormendy2016}. Within the catalogue of 175 galaxies there are 12 face-on galaxies with \textit{i} $< 30\degr$, 22 galaxies with \textit{i} $< 35\degr$, and 28 galaxies with \textit{i} $< 40\degr$. If the 12 face-on galaxies with \textit{i} $< 30\degr$ are removed from the database, $\sigma$ decreases 0.006 and rms decreases 0.006. 
   
   For 3082 of 3391 data points in the catalogue, circular velocity uncertainty is $\leq$ 15$\%$. Improving the uncertainty to $\leq$ 10$\%$ reduces the data points to 2807; improving the uncertainty to $\leq$ 5$\%$ reduces the data points to 1967. 

To maximise the data set and minimise assumptions, the present study excludes none of the 175 galaxies in the catalogue and removes none of the 3391 data points.

\section{Results} \label{conpilation}

\begin{figure}
\begin{center}
\includegraphics[width=8cm]{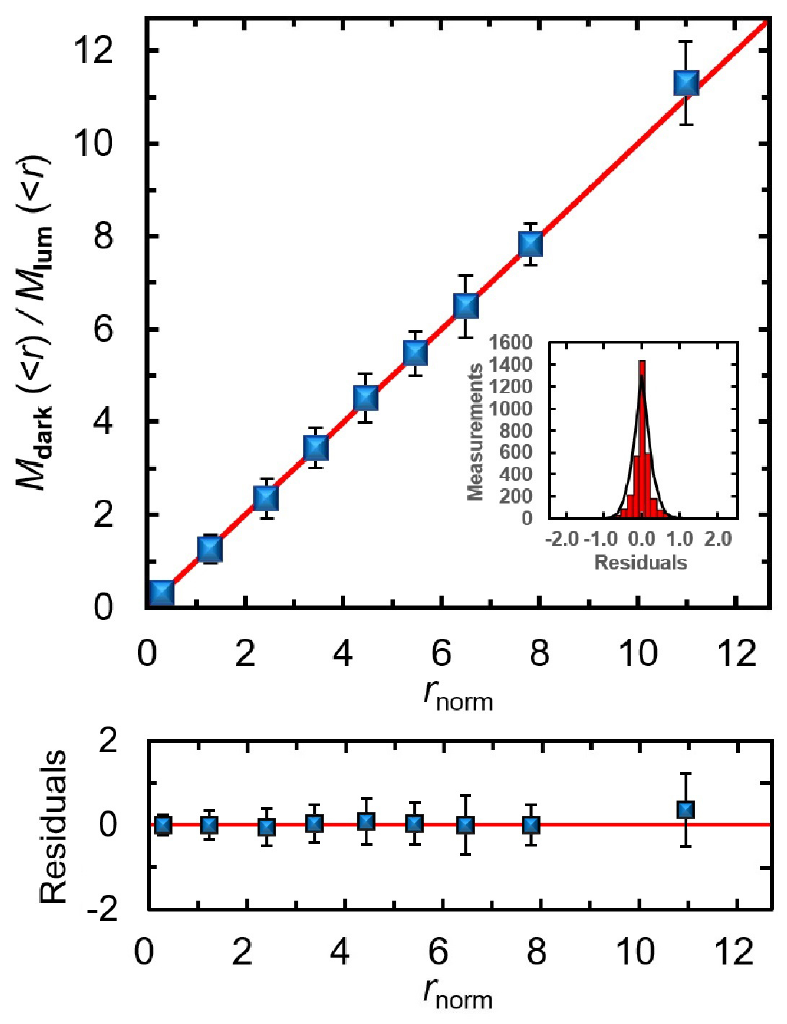}
\end{center}
\caption{Ratio of enclosed dark to luminous mass as a function of normalised radius for the complete set of 175 SPARC galaxies. In upper panel data for 3391 points are binned at approximate intervals of 1.0 $\textit{r}_{norm}$, with the last bin extended to include outliers. Filled squares denote the mean of binned data. Error bars represent 1$\sigma$ of data in each bin. The error bar for 1$\sigma$ =  0.21 in the first bin is not visible. Solid red line corresponds to a linear scaling of enclosed dark to luminous mass. Inset shows the histogram of all residuals and a Gaussian of width $\sigma$ = 0.31. Lower panel displays the residuals as a function of $\textit{r}_{norm}$. Error bars on the binned data represent 1$\sigma$.    \label{fig:dmfig3}}
\end{figure}

To evaluate the scaling of enclosed dark to luminous mass in a broad range of galaxies, we examine the full set of 175 galaxies in the SPARC catalogue. Figure~\ref{fig:dmfig3} shows the scaling of enclosed dark to luminous mass as a function of normalised radius for the full data set. The 3391 data points are binned at approximate intervals of 1.0 $\textit{r}_{norm}$, with the last bin extended to include outliers. Filled squares denote the mean of binned data. Error bars represent 1$\sigma$ of data in each bin. The error bar for 1$\sigma$ = 0.21 in the first bin is not visible. The solid red line corresponds to a linear relation. The inset shows the histogram of all residuals and a Gaussian of width $\sigma$ = 0.31. The lower panel displays the residuals as a function of $\textit{r}_{norm}$. Error bars on the binned data represent 1$\sigma$. Residuals of each galaxy are the same prior to and after normalising.

The complete set of 3391 data points spanning 175 galaxies evinces a linear scaling of enclosed dark to luminous mass. Uncertainties are quantified by rms and $\sigma$, with rms = 0.31 and $\sigma$ = 0.31. For all 3391 data points, the correlation coefficient r = 0.9838.

  In the outer region of disc galaxies the radial mass profile of the dark matter scales linearly with the luminous mass to commonly produce a flat rotation curve. In the SPARC catalogue, 40 of the 175 galaxies do not exhibit a flat rotation curve. Whether producing a flat or non-flat rotation curve, the distribution of dark matter particles in the studied galaxies results in a linear scaling of enclosed dark to luminous mass that extends from the inner region to the outermost data point. Such linear scaling is not anticipated by standard hierarchical galaxy formation theory: the merger of small haloes into larger haloes is designed to explain structure rather than to produce a linear relation; accordingly a linear scaling of enclosed dark to luminous mass has not been reported as an outcome of any galaxy formation model. It is uncertain what would produce this linear relation. Baryonic processes are challenged to account for the linear scaling, and no dark matter candidate possesses a theoretical property that would result in a linear relation. 
 
 The empirical linear relation may be used to test and constrain galaxy formation models using the closeness of the observed relation as a benchmark. 

The linearity of enclosed dark to luminous mass may indicate the existence of new dark matter candidates, or the presence of an astrophysical process beyond standard galaxy formation cosmology.  

\section{Summary} \label{summary}

Standard cosmological theory posits that galaxies formed in the early universe within the potential wells of dark matter haloes. Small haloes merged with larger haloes to form the galaxies, groups, and clusters of the present universe. In standard cosmology, dark matter haloes are comprised of theoretical particles that neither emit, nor absorb, nor reflect radiation at any wavelength. Well motivated dark matter candidates include WIMPs, warm dark matter, self-interacting dark matter, axions, and primordial black holes. 

Measurements from 175 galaxies spanning a broad range of morphology reveal a linear scaling of enclosed dark to luminous mass that is not anticipated by standard galaxy formation cosmology. In the framework of standard galaxy formation theory, the linear scaling of enclosed dark to luminous mass would require tuning the dark matter profile of each galaxy. Such tuning would be required whether the dark matter consisted of WIMP dark matter, warm dark matter, primordial black holes, axions, self-interacting dark matter, or a combination of dark matter candidates. 

In the studied galaxies, wherever dark mass is found--throughout a spherical E0 galaxy, within the arms of spiral galaxies, in the ill-defined structure of irregular galaxies--the dark matter is distributed in such a way as to produce a linear scaling of enclosed dark to luminous mass. No dark matter candidate possesses a theoretical property that would result in a linear relation. Further work will lead to dark matter properties or astrophysical processes that could produce the close linear relation.

\section*{Acknowledgements}

The author thanks John Kielkopf for comments and discussions, Federico Lelli for comments on a draft of the paper as well as observations regarding the SPARC database, and the reviewer for useful comments, This work made use of the SPARC (Spitzer Photometry \& Accurate Rotation Curves) catalogue.

\section*{Data Availability}

The data underlying this article are available in the Zenodo repository, at https://doi.org/10.5281/zenodo.6416252.



\bibliographystyle{mnras}
\bibliography{linearity} 

\begin{thebibliography}{}
\makeatletter
\relax
\def\mn@urlcharsother{\let\do\@makeother \do\$\do\&\do\#\do\^\do\_\do\%\do\~}
\def\mn@doi{\begingroup\mn@urlcharsother \@ifnextchar [ {\mn@doi@}
  {\mn@doi@[]}}
\def\mn@doi@[#1]#2{\def\@tempa{#1}\ifx\@tempa\@empty \href
  {http://dx.doi.org/#2} {doi:#2}\else \href {http://dx.doi.org/#2} {#1}\fi
  \endgroup}
\def\mn@eprint#1#2{\mn@eprint@#1:#2::\@nil}
\def\mn@eprint@arXiv#1{\href {http://arxiv.org/abs/#1} {{\tt arXiv:#1}}}
\def\mn@eprint@dblp#1{\href {http://dblp.uni-trier.de/rec/bibtex/#1.xml}
  {dblp:#1}}
\def\mn@eprint@#1:#2:#3:#4\@nil{\def\@tempa {#1}\def\@tempb {#2}\def\@tempc
  {#3}\ifx \@tempc \@empty \let \@tempc \@tempb \let \@tempb \@tempa \fi \ifx
  \@tempb \@empty \def\@tempb {arXiv}\fi \@ifundefined
  {mn@eprint@\@tempb}{\@tempb:\@tempc}{\expandafter \expandafter \csname
  mn@eprint@\@tempb\endcsname \expandafter{\@tempc}}}

\bibitem[\protect\citeauthoryear{{Amorisco}, {Agnello}  \& {Evans}}{{Amorisco}
  et~al.}{2013}]{amorisco2013}
{Amorisco} N.~C.,  {Agnello} A.,   {Evans} N.~W.,  2013, \mn@doi [\mnras]
  {10.1093/mnrasl/sls031}, \href
  {https://ui.adsabs.harvard.edu/abs/2013MNRAS.429L..89A} {429, L89}

\bibitem[\protect\citeauthoryear{{Basu}, {Roy}, {Choudhuri}, {Datta}  \&
  {Sarkar}}{{Basu} et~al.}{2021}]{Basu2021}
{Basu} A.,  {Roy} N.,  {Choudhuri} S.,  {Datta} K.~K.,   {Sarkar} D.,  2021,
  \mn@doi [\mnras] {10.1093/mnras/stab120}, \href
  {https://ui.adsabs.harvard.edu/abs/2021MNRAS.502.1605B} {502, 1605}

\bibitem[\protect\citeauthoryear{{Bauer}, {Marsh}, {Hlo{\v{z}}ek},
  {Padmanabhan}  \& {Lagu{\"e}}}{{Bauer} et~al.}{2021}]{Bauer2021}
{Bauer} J.~B.,  {Marsh} D. J.~E.,  {Hlo{\v{z}}ek} R.,  {Padmanabhan} H.,
  {Lagu{\"e}} A.,  2021, \mn@doi [\mnras] {10.1093/mnras/staa3300}, \href
  {https://ui.adsabs.harvard.edu/abs/2021MNRAS.500.3162B} {500, 3162}

\bibitem[\protect\citeauthoryear{{Bekki} \& {Stanimirovi{\'c}}}{{Bekki} \&
  {Stanimirovi{\'c}}}{2009}]{bekki2009}
{Bekki} K.,  {Stanimirovi{\'c}} S.,  2009, \mn@doi [\mnras]
  {10.1111/j.1365-2966.2009.14514.x}, \href
  {https://ui.adsabs.harvard.edu/abs/2009MNRAS.395..342B} {395, 342}

\bibitem[\protect\citeauthoryear{{Bernal}, {Fern{\'a}ndez-Hern{\'a}ndez},
  {Matos}  \& {Rodr{\'\i}guez-Meza}}{{Bernal} et~al.}{2018}]{Bernal2018}
{Bernal} T.,  {Fern{\'a}ndez-Hern{\'a}ndez} L.~M.,  {Matos} T.,
  {Rodr{\'\i}guez-Meza} M.~A.,  2018, \mn@doi [\mnras] {10.1093/mnras/stx3208},
  \href {https://ui.adsabs.harvard.edu/abs/2018MNRAS.475.1447B} {475, 1447}

\bibitem[\protect\citeauthoryear{{Binney} \& {Evans}}{{Binney} \&
  {Evans}}{2001}]{binney2001}
{Binney} J.~J.,  {Evans} N.~W.,  2001, \mn@doi [\mnras]
  {10.1046/j.1365-8711.2001.04968.x}, \href
  {https://ui.adsabs.harvard.edu/abs/2001MNRAS.327L..27B} {327, L27}

\bibitem[\protect\citeauthoryear{{Blumenthal}, {Faber}, {Primack}  \&
  {Rees}}{{Blumenthal} et~al.}{1984}]{Blumenthal1984}
{Blumenthal} G.~R.,  {Faber} S.~M.,  {Primack} J.~R.,   {Rees} M.~J.,  1984,
  \mn@doi [\nat] {10.1038/311517a0}, \href
  {https://ui.adsabs.harvard.edu/abs/1984Natur.311..517B} {311, 517}

\bibitem[\protect\citeauthoryear{{Boylan-Kolchin}, {Bullock}  \&
  {Kaplinghat}}{{Boylan-Kolchin} et~al.}{2011}]{Boylan-Kolchin2011}
{Boylan-Kolchin} M.,  {Bullock} J.~S.,   {Kaplinghat} M.,  2011, \mn@doi
  [\mnras] {10.1111/j.1745-3933.2011.01074.x}, \href
  {https://ui.adsabs.harvard.edu/abs/2011MNRAS.415L..40B} {415, L40}

\bibitem[\protect\citeauthoryear{{Bromm}, {Yoshida}, {Hernquist}  \&
  {McKee}}{{Bromm} et~al.}{2009}]{Bromm2009}
{Bromm} V.,  {Yoshida} N.,  {Hernquist} L.,   {McKee} C.~F.,  2009, \mn@doi
  [\nat] {10.1038/nature07990}, \href
  {https://ui.adsabs.harvard.edu/abs/2009Natur.459...49B} {459, 49}

\bibitem[\protect\citeauthoryear{{Bullock} \& {Boylan-Kolchin}}{{Bullock} \&
  {Boylan-Kolchin}}{2017}]{bullock2017}
{Bullock} J.~S.,  {Boylan-Kolchin} M.,  2017, \mn@doi [\araa]
  {10.1146/annurev-astro-091916-055313}, \href
  {https://ui.adsabs.harvard.edu/abs/2017ARA&A..55..343B} {55, 343}

\bibitem[\protect\citeauthoryear{{Carr} \& {Silk}}{{Carr} \&
  {Silk}}{2018}]{Carr2018}
{Carr} B.,  {Silk} J.,  2018, \mn@doi [\mnras] {10.1093/mnras/sty1204}, \href
  {https://ui.adsabs.harvard.edu/abs/2018MNRAS.478.3756C} {478, 3756}

\bibitem[\protect\citeauthoryear{{Chan}, {Kere{\v{s}}}, {O{\~n}orbe},
  {Hopkins}, {Muratov}, {Faucher-Gigu{\`e}re}  \& {Quataert}}{{Chan}
  et~al.}{2015}]{chan2015}
{Chan} T.~K.,  {Kere{\v{s}}} D.,  {O{\~n}orbe} J.,  {Hopkins} P.~F.,  {Muratov}
  A.~L.,  {Faucher-Gigu{\`e}re} C.~A.,   {Quataert} E.,  2015, \mn@doi [\mnras]
  {10.1093/mnras/stv2165}, \href
  {https://ui.adsabs.harvard.edu/abs/2015MNRAS.454.2981C} {454, 2981}

\bibitem[\protect\citeauthoryear{{Chatterjee}, {Dayal}, {Choudhury}  \&
  {Hutter}}{{Chatterjee} et~al.}{2019}]{Chatterjee2019}
{Chatterjee} A.,  {Dayal} P.,  {Choudhury} T.~R.,   {Hutter} A.,  2019, \mn@doi
  [\mnras] {10.1093/mnras/stz1444}, \href
  {https://ui.adsabs.harvard.edu/abs/2019MNRAS.487.3560C} {487, 3560}

\bibitem[\protect\citeauthoryear{{Clowe}, {Brada{\v{c}}}, {Gonzalez},
  {Markevitch}, {Randall}, {Jones}  \& {Zaritsky}}{{Clowe}
  et~al.}{2006}]{Clowe2006}
{Clowe} D.,  {Brada{\v{c}}} M.,  {Gonzalez} A.~H.,  {Markevitch} M.,  {Randall}
  S.~W.,  {Jones} C.,   {Zaritsky} D.,  2006, \mn@doi [\apjl] {10.1086/508162},
  \href {https://ui.adsabs.harvard.edu/abs/2006ApJ...648L.109C} {648, L109}

\bibitem[\protect\citeauthoryear{{Cook} et~al.,}{{Cook}
  et~al.}{2020}]{Cook2020}
{Cook} R.~H.~W.,  et~al., 2020, \mn@doi [\mnras] {10.1093/mnras/staa726}, \href
  {https://ui.adsabs.harvard.edu/abs/2020MNRAS.494..135C} {494, 135}

\bibitem[\protect\citeauthoryear{{Correa}}{{Correa}}{2021}]{Correa2021}
{Correa} C.~A.,  2021, \mn@doi [\mnras] {10.1093/mnras/stab506}, \href
  {https://ui.adsabs.harvard.edu/abs/2021MNRAS.503..920C} {503, 920}

\bibitem[\protect\citeauthoryear{{Dashyan}, {Silk}, {Mamon}, {Dubois}  \&
  {Hartwig}}{{Dashyan} et~al.}{2018}]{dashyan2018}
{Dashyan} G.,  {Silk} J.,  {Mamon} G.~A.,  {Dubois} Y.,   {Hartwig} T.,  2018,
  \mn@doi [\mnras] {10.1093/mnras/stx2716}, \href
  {https://ui.adsabs.harvard.edu/abs/2018MNRAS.473.5698D} {473, 5698}

\bibitem[\protect\citeauthoryear{{De Blok}, {McGaugh}, {Bosma}  \& {Rubin}}{{De
  Blok} et~al.}{2001}]{deblok2001}
{De Blok} W.~J.~G.,  {McGaugh} S.~S.,  {Bosma} A.,   {Rubin} V.~C.,  2001,
  \mn@doi [\apjl] {10.1086/320262}, \href
  {https://ui.adsabs.harvard.edu/abs/2001ApJ...552L..23D} {552, L23}

\bibitem[\protect\citeauthoryear{{Despali}, {Sparre}, {Vegetti},
  {Vogelsberger}, {Zavala}  \& {Marinacci}}{{Despali}
  et~al.}{2019}]{Despali2019}
{Despali} G.,  {Sparre} M.,  {Vegetti} S.,  {Vogelsberger} M.,  {Zavala} J.,
  {Marinacci} F.,  2019, \mn@doi [\mnras] {10.1093/mnras/stz273}, \href
  {https://ui.adsabs.harvard.edu/abs/2019MNRAS.484.4563D} {484, 4563}

\bibitem[\protect\citeauthoryear{{Di Cintio}, {Knebe}, {Libeskind}, {Yepes},
  {Gottl{\"o}ber}  \& {Hoffman}}{{Di Cintio} et~al.}{2011}]{dicintio2011}
{Di Cintio} A.,  {Knebe} A.,  {Libeskind} N.~I.,  {Yepes} G.,  {Gottl{\"o}ber}
  S.,   {Hoffman} Y.,  2011, \mn@doi [\mnras]
  {10.1111/j.1745-3933.2011.01123.x}, \href
  {https://ui.adsabs.harvard.edu/abs/2011MNRAS.417L..74D} {417, L74}

\bibitem[\protect\citeauthoryear{Freeman}{Freeman}{2003}]{Freeman2003}
Freeman K.~C.,  2003, Sci, 302, 1902

\bibitem[\protect\citeauthoryear{{Galanti}, {Roncadelli}, {De Angelis}  \&
  {Bignami}}{{Galanti} et~al.}{2020}]{Galanti2020}
{Galanti} G.,  {Roncadelli} M.,  {De Angelis} A.,   {Bignami} G.~F.,  2020,
  \mn@doi [\mnras] {10.1093/mnras/stz3410}, \href
  {https://ui.adsabs.harvard.edu/abs/2020MNRAS.493.1553G} {493, 1553}

\bibitem[\protect\citeauthoryear{{Garrison-Kimmel}, {Boylan-Kolchin}, {Bullock}
   \& {Kirby}}{{Garrison-Kimmel} et~al.}{2014}]{garrison-kimmel2014}
{Garrison-Kimmel} S.,  {Boylan-Kolchin} M.,  {Bullock} J.~S.,   {Kirby} E.~N.,
  2014, \mn@doi [\mnras] {10.1093/mnras/stu1477}, \href
  {https://ui.adsabs.harvard.edu/abs/2014MNRAS.444..222G} {444, 222}

\bibitem[\protect\citeauthoryear{{Garzilli}, {Magalich}, {Ruchayskiy}  \&
  {Boyarsky}}{{Garzilli} et~al.}{2021}]{Garzilli2021}
{Garzilli} A.,  {Magalich} A.,  {Ruchayskiy} O.,   {Boyarsky} A.,  2021,
  \mn@doi [\mnras] {10.1093/mnras/stab192}, \href
  {https://ui.adsabs.harvard.edu/abs/2021MNRAS.502.2356G} {502, 2356}

\bibitem[\protect\citeauthoryear{{Gentile}, {Salucci}, {Klein}, {Vergani}  \&
  {Kalberla}}{{Gentile} et~al.}{2004}]{Gentile2004}
{Gentile} G.,  {Salucci} P.,  {Klein} U.,  {Vergani} D.,   {Kalberla} P.,
  2004, \mn@doi [\mnras] {10.1111/j.1365-2966.2004.07836.x}, \href
  {https://ui.adsabs.harvard.edu/abs/2004MNRAS.351..903G} {351, 903}

\bibitem[\protect\citeauthoryear{{Gentile}, {Salucci}, {Klein}  \&
  {Granato}}{{Gentile} et~al.}{2007}]{gentile2007}
{Gentile} G.,  {Salucci} P.,  {Klein} U.,   {Granato} G.~L.,  2007, \mn@doi
  [\mnras] {10.1111/j.1365-2966.2006.11283.x}, \href
  {https://ui.adsabs.harvard.edu/abs/2007MNRAS.375..199G} {375, 199}

\bibitem[\protect\citeauthoryear{{Giar{\'e}}, {Di Valentino}, {Melchiorri}  \&
  {Mena}}{{Giar{\'e}} et~al.}{2021}]{Giare2021}
{Giar{\'e}} W.,  {Di Valentino} E.,  {Melchiorri} A.,   {Mena} O.,  2021,
  \mn@doi [\mnras] {10.1093/mnras/stab1442}, \href
  {https://ui.adsabs.harvard.edu/abs/2021MNRAS.505.2703G} {505, 2703}

\bibitem[\protect\citeauthoryear{{Gilman}, {Birrer}, {Nierenberg}, {Treu}, {Du}
   \& {Benson}}{{Gilman} et~al.}{2020}]{Gilman2020}
{Gilman} D.,  {Birrer} S.,  {Nierenberg} A.,  {Treu} T.,  {Du} X.,   {Benson}
  A.,  2020, \mn@doi [\mnras] {10.1093/mnras/stz3480}, \href
  {https://ui.adsabs.harvard.edu/abs/2020MNRAS.491.6077G} {491, 6077}

\bibitem[\protect\citeauthoryear{{Governato} et~al.,}{{Governato}
  et~al.}{2012}]{governato2012}
{Governato} F.,  et~al., 2012, \mn@doi [\mnras]
  {10.1111/j.1365-2966.2012.20696.x}, \href
  {https://ui.adsabs.harvard.edu/abs/2012MNRAS.422.1231G} {422, 1231}

\bibitem[\protect\citeauthoryear{{Hallenbeck} et~al.,}{{Hallenbeck}
  et~al.}{2014}]{Hallenbeck2014}
{Hallenbeck} G.,  et~al., 2014, \mn@doi [\aj] {10.1088/0004-6256/148/4/69},
  \href {https://ui.adsabs.harvard.edu/abs/2014AJ....148...69H} {148, 69}

\bibitem[\protect\citeauthoryear{{Hermans}, {Banik}, {Weniger}, {Bertone}  \&
  {Louppe}}{{Hermans} et~al.}{2021}]{Hermans2021}
{Hermans} J.,  {Banik} N.,  {Weniger} C.,  {Bertone} G.,   {Louppe} G.,  2021,
  \mn@doi [\mnras] {10.1093/mnras/stab2181}, \href
  {https://ui.adsabs.harvard.edu/abs/2021MNRAS.507.1999H} {507, 1999}

\bibitem[\protect\citeauthoryear{{Iwanus}, {Elahi}, {List}  \&
  {Lewis}}{{Iwanus} et~al.}{2019}]{Iwanus2019}
{Iwanus} N.,  {Elahi} P.~J.,  {List} F.,   {Lewis} G.~F.,  2019, \mn@doi
  [\mnras] {10.1093/mnras/stz435}, \href
  {https://ui.adsabs.harvard.edu/abs/2019MNRAS.485.1420I} {485, 1420}

\bibitem[\protect\citeauthoryear{{Keeton}}{{Keeton}}{2001}]{keeton2001}
{Keeton} C.~R.,  2001, \mn@doi [\apj] {10.1086/323237}, \href
  {https://ui.adsabs.harvard.edu/abs/2001ApJ...561...46K} {561, 46}

\bibitem[\protect\citeauthoryear{{Kim}}{{Kim}}{2021}]{Kim2021}
{Kim} H.,  2021, \mn@doi [\mnras] {10.1093/mnras/stab1222}, \href
  {https://ui.adsabs.harvard.edu/abs/2021MNRAS.504.5475K} {504, 5475}

\bibitem[\protect\citeauthoryear{{Kormendy} \& {Freeman}}{{Kormendy} \&
  {Freeman}}{2016}]{Kormendy2016}
{Kormendy} J.,  {Freeman} K.~C.,  2016, \mn@doi [\apj]
  {10.3847/0004-637X/817/2/84}, \href
  {http://adsabs.harvard.edu/abs/2016ApJ...817...84K} {817, 84}

\bibitem[\protect\citeauthoryear{{Lelli}, {McGaugh}  \& {Schombert}}{{Lelli}
  et~al.}{2016}]{Lelli2016}
{Lelli} F.,  {McGaugh} S.~S.,   {Schombert} J.~M.,  2016, \mn@doi [\aj]
  {10.3847/0004-6256/152/6/157}, \href
  {https://ui.adsabs.harvard.edu/abs/2016AJ....152..157L} {152, 157}

\bibitem[\protect\citeauthoryear{{Leong}, {Schive}, {Zhang}  \&
  {Chiueh}}{{Leong} et~al.}{2019}]{Leong2019}
{Leong} K.-H.,  {Schive} H.-Y.,  {Zhang} U.-H.,   {Chiueh} T.,  2019, \mn@doi
  [\mnras] {10.1093/mnras/stz271}, \href
  {https://ui.adsabs.harvard.edu/abs/2019MNRAS.484.4273L} {484, 4273}

\bibitem[\protect\citeauthoryear{{Lovas} \& {Kielkopf}}{{Lovas} \&
  {Kielkopf}}{2014}]{Lovas2014}
{Lovas} S.,  {Kielkopf} J.~F.,  2014, \mn@doi [\aj]
  {10.1088/0004-6256/147/6/135}, \href
  {http://adsabs.harvard.edu/abs/2014AJ....147..135L} {147, 135}

\bibitem[\protect\citeauthoryear{{Madau}, {Shen}  \& {Governato}}{{Madau}
  et~al.}{2014}]{madau2014}
{Madau} P.,  {Shen} S.,   {Governato} F.,  2014, \mn@doi [\apjl]
  {10.1088/2041-8205/789/1/L17}, \href
  {https://ui.adsabs.harvard.edu/abs/2014ApJ...789L..17M} {789, L17}

\bibitem[\protect\citeauthoryear{{Meidt} et~al.,}{{Meidt}
  et~al.}{2014}]{Meidt2014}
{Meidt} S.~E.,  et~al., 2014, \mn@doi [\apj] {10.1088/0004-637X/788/2/144},
  \href {http://adsabs.harvard.edu/abs/2014ApJ...788..144M} {788, 144}

\bibitem[\protect\citeauthoryear{{Milgrom}}{{Milgrom}}{1983}]{Milgrom1983}
{Milgrom} M.,  1983, \apj, \href {xx} {270, 365}

\bibitem[\protect\citeauthoryear{{Moore}, {Governato}, {Quinn}, {Stadel}  \&
  {Lake}}{{Moore} et~al.}{1998}]{moore1998}
{Moore} B.,  {Governato} F.,  {Quinn} T.,  {Stadel} J.,   {Lake} G.,  1998,
  \mn@doi [\apjl] {10.1086/311333}, \href
  {https://ui.adsabs.harvard.edu/abs/1998ApJ...499L...5M} {499, L5}

\bibitem[\protect\citeauthoryear{{Naik}, {Puchwein}, {Davis}, {Sijacki}  \&
  {Desmond}}{{Naik} et~al.}{2019}]{Naik2019}
{Naik} A.~P.,  {Puchwein} E.,  {Davis} A.-C.,  {Sijacki} D.,   {Desmond} H.,
  2019, \mn@doi [\mnras] {10.1093/mnras/stz2131}, \href
  {https://ui.adsabs.harvard.edu/abs/2019MNRAS.489..771N} {489, 771}

\bibitem[\protect\citeauthoryear{{Navarro}, {Frenk}  \& {White}}{{Navarro}
  et~al.}{1996}]{navarro1996}
{Navarro} J.~F.,  {Frenk} C.~S.,   {White} S. D.~M.,  1996, \mn@doi [\apj]
  {10.1086/177173}, \href
  {https://ui.adsabs.harvard.edu/abs/1996ApJ...462..563N} {462, 563}

\bibitem[\protect\citeauthoryear{{Pointecouteau} \& {Silk}}{{Pointecouteau} \&
  {Silk}}{2005}]{Silk2005}
{Pointecouteau} E.,  {Silk} J.,  2005, \mnras, \href {xx} {364, 654}

\bibitem[\protect\citeauthoryear{{Ponomareva} et~al.,}{{Ponomareva}
  et~al.}{2021}]{Ponomareva2021}
{Ponomareva} A.~A.,  et~al., 2021, \mn@doi [\mnras] {10.1093/mnras/stab2654},
  \href {https://ui.adsabs.harvard.edu/abs/2021MNRAS.508.1195P} {508, 1195}

\bibitem[\protect\citeauthoryear{{Robertson}, {Massey}, {Eke}, {Schaye}  \&
  {Theuns}}{{Robertson} et~al.}{2021}]{Robertson2021}
{Robertson} A.,  {Massey} R.,  {Eke} V.,  {Schaye} J.,   {Theuns} T.,  2021,
  \mn@doi [\mnras] {10.1093/mnras/staa3954}, \href
  {https://ui.adsabs.harvard.edu/abs/2021MNRAS.501.4610R} {501, 4610}

\bibitem[\protect\citeauthoryear{{Robles}, {Kelley}, {Bullock}  \&
  {Kaplinghat}}{{Robles} et~al.}{2019}]{Robles2019}
{Robles} V.~H.,  {Kelley} T.,  {Bullock} J.~S.,   {Kaplinghat} M.,  2019,
  \mn@doi [\mnras] {10.1093/mnras/stz2345}, \href
  {https://ui.adsabs.harvard.edu/abs/2019MNRAS.490.2117R} {490, 2117}

\bibitem[\protect\citeauthoryear{{Rodrigues}, {del Popolo}, {Marra}  \& {de
  Oliveira}}{{Rodrigues} et~al.}{2017}]{rodrigues2017}
{Rodrigues} D.~C.,  {del Popolo} A.,  {Marra} V.,   {de Oliveira} P. L.~C.,
  2017, \mn@doi [\mnras] {10.1093/mnras/stx1384}, \href
  {https://ui.adsabs.harvard.edu/abs/2017MNRAS.470.2410R} {470, 2410}

\bibitem[\protect\citeauthoryear{{S{\'a}nchez-Salcedo}, {Hidalgo-G{\'a}mez}  \&
  {Mart{\'{\i}}nez-Garc{\'{\i}}a}}{{S{\'a}nchez-Salcedo}
  et~al.}{2013}]{Sanchez-Salcedo2013}
{S{\'a}nchez-Salcedo} F.~J.,  {Hidalgo-G{\'a}mez} A.~M.,
  {Mart{\'{\i}}nez-Garc{\'{\i}}a} E.~E.,  2013, \mn@doi [\aj]
  {10.1088/0004-6256/145/3/61}, \href
  {http://adsabs.harvard.edu/abs/2013AJ....145...61S} {145, 61}

\bibitem[\protect\citeauthoryear{{Santos-Santos}, {Di Cintio}, {Brook},
  {Macci{\`o}}, {Dutton}  \& {Dom{\'\i}nguez-Tenreiro}}{{Santos-Santos}
  et~al.}{2018}]{Santos2018}
{Santos-Santos} I.~M.,  {Di Cintio} A.,  {Brook} C.~B.,  {Macci{\`o}} A.,
  {Dutton} A.,   {Dom{\'\i}nguez-Tenreiro} R.,  2018, \mn@doi [\mnras]
  {10.1093/mnras/stx2660}, \href
  {https://ui.adsabs.harvard.edu/abs/2018MNRAS.473.4392S} {473, 4392}

\bibitem[\protect\citeauthoryear{{Schuberth}, {Richtler}, {Dirsch}, {Hilker},
  {Larsen}, {Kissler-Patig}  \& {Mebold}}{{Schuberth}
  et~al.}{2006}]{Schuberth2006}
{Schuberth} Y.,  {Richtler} T.,  {Dirsch} B.,  {Hilker} M.,  {Larsen} S.~S.,
  {Kissler-Patig} M.,   {Mebold} U.,  2006, \mn@doi [\aap]
  {10.1051/0004-6361:20053134}, \href
  {https://ui.adsabs.harvard.edu/abs/2006A&A...459..391S} {459, 391}

\bibitem[\protect\citeauthoryear{{Spano}, {Marcelin}, {Amram}, {Carignan},
  {Epinat}  \& {Hernandez}}{{Spano} et~al.}{2008}]{spano2008}
{Spano} M.,  {Marcelin} M.,  {Amram} P.,  {Carignan} C.,  {Epinat} B.,
  {Hernandez} O.,  2008, \mn@doi [\mnras] {10.1111/j.1365-2966.2007.12545.x},
  \href {https://ui.adsabs.harvard.edu/abs/2008MNRAS.383..297S} {383, 297}

\bibitem[\protect\citeauthoryear{{Stegmann}, {Capelo}, {Bortolas}  \&
  {Mayer}}{{Stegmann} et~al.}{2020}]{Stegmann2020}
{Stegmann} J.,  {Capelo} P.~R.,  {Bortolas} E.,   {Mayer} L.,  2020, \mn@doi
  [\mnras] {10.1093/mnras/staa170}, \href
  {https://ui.adsabs.harvard.edu/abs/2020MNRAS.492.5247S} {492, 5247}

\bibitem[\protect\citeauthoryear{{Sureda}, {Maga{\~n}a}, {Araya}  \&
  {Padilla}}{{Sureda} et~al.}{2021}]{Sureda2021}
{Sureda} J.,  {Maga{\~n}a} J.,  {Araya} I.~J.,   {Padilla} N.~D.,  2021,
  \mn@doi [\mnras] {10.1093/mnras/stab2450}, \href
  {https://ui.adsabs.harvard.edu/abs/2021MNRAS.507.4804S} {507, 4804}

\bibitem[\protect\citeauthoryear{{Thorpe-Morgan}, {Malyshev}, {Stegen},
  {Santangelo}  \& {Jochum}}{{Thorpe-Morgan} et~al.}{2021}]{Thorpe-Morgan2021}
{Thorpe-Morgan} C.,  {Malyshev} D.,  {Stegen} C.-A.,  {Santangelo} A.,
  {Jochum} J.,  2021, \mn@doi [\mnras] {10.1093/mnras/stab208}, \href
  {https://ui.adsabs.harvard.edu/abs/2021MNRAS.502.4039T} {502, 4039}

\bibitem[\protect\citeauthoryear{{Tortora}, {Posti}, {Koopmans}  \&
  {Napolitano}}{{Tortora} et~al.}{2019}]{Tortora2019}
{Tortora} C.,  {Posti} L.,  {Koopmans} L.~V.~E.,   {Napolitano} N.~R.,  2019,
  \mn@doi [\mnras] {10.1093/mnras/stz2320}, \href
  {https://ui.adsabs.harvard.edu/abs/2019MNRAS.489.5483T} {489, 5483}

\bibitem[\protect\citeauthoryear{{Wetzel}, {Hopkins}, {Kim},
  {Faucher-Gigu{\`e}re}, {Kere{\v{s}}}  \& {Quataert}}{{Wetzel}
  et~al.}{2016}]{wetzel2016}
{Wetzel} A.~R.,  {Hopkins} P.~F.,  {Kim} J.-h.,  {Faucher-Gigu{\`e}re} C.-A.,
  {Kere{\v{s}}} D.,   {Quataert} E.,  2016, \mn@doi [\apjl]
  {10.3847/2041-8205/827/2/L23}, \href
  {https://ui.adsabs.harvard.edu/abs/2016ApJ...827L..23W} {827, L23}

\bibitem[\protect\citeauthoryear{{White} \& {Rees}}{{White} \&
  {Rees}}{1978}]{White1978}
{White} S.~D.~M.,  {Rees} M.~J.,  1978, \mn@doi [\mnras]
  {10.1093/mnras/183.3.341}, \href
  {https://ui.adsabs.harvard.edu/abs/1978MNRAS.183..341W} {183, 341}

\bibitem[\protect\citeauthoryear{{Zolotov} et~al.,}{{Zolotov}
  et~al.}{2012}]{zolotov2012}
{Zolotov} A.,  et~al., 2012, \mn@doi [\apj] {10.1088/0004-637X/761/1/71}, \href
  {https://ui.adsabs.harvard.edu/abs/2012ApJ...761...71Z} {761, 71}

\makeatother
\end{thebibliography}





\bsp	
\label{lastpage}
\end{document}